# First results of the experiment to search for 2β decay of $^{106}$Cd with $^{106}$CdWO$_4$ crystal scintillator in coincidence with four crystals HPGe detector[a]

V.I. Tretyak[1,b], P. Belli[2], R. Bernabei[2,3], V.B. Brudanin[4], F. Cappella[5,6], V. Caracciolo[7], R. Cerulli[7], D.M. Chernyak[1], F.A. Danevich[1], S. D'Angelo[2,3], A. Incicchitti[5], M. Laubenstein[7], V.M. Mokina[1], D.V. Poda[1], O.G. Polischuk[1,6], R.B. Podviyanuk[1], I.A. Tupitsyna[8]

[1] Institute for Nuclear Research, MSP 03680 Kyiv, Ukraine
[2] Dipartimento di Fisica, Università di Roma "Tor Vergata", I-00133 Rome, Italy
[3] INFN sezione Roma "Tor Vergata", I-00133 Rome, Italy
[4] Joint Institute for Nuclear Research, 141980 Dubna, Russia
[5] Dipartimento di Fisica, Università di Roma "La Sapienza", I-00185 Rome, Italy
[6] INFN, sezione di Roma, I-00185 Rome, Italy
[7] INFN, Laboratori Nazionali del Gran Sasso, I-67100 Assergi (AQ), Italy
[8] Institute of Scintillation Materials, 61001 Kharkiv, Ukraine

**Abstract.** An experiment to search for double beta processes in $^{106}$Cd by using cadmium tungstate crystal scintillator enriched in $^{106}$Cd ($^{106}$CdWO$_4$) in coincidence with the four crystals HPGe detector GeMulti is in progress at the STELLA facility of the Gran Sasso underground laboratory of INFN (Italy). The $^{106}$CdWO$_4$ scintillator is viewed by a low-background photomultiplier tube through a lead tungstate crystal light-guide produced from deeply purified archaeological lead to suppress γ quanta from the photomultiplier tube. Here we report the first results of the experiment after 3233 hours of the data taking. A few new improved limits on double beta processes in $^{106}$Cd are obtained, in particular $T_{1/2}^{2\nu\varepsilon\beta+} \geq 8.4 \times 10^{20}$ yr at 90% C.L.

## 1 Introduction

Experiments to search for double beta (2β) decay are considered as a promising way to investigate properties of neutrino and search for effects beyond the Standard Model of particles [1, 2, 3, 4, 5]. The isotope $^{106}$Cd (energy of decay $Q_{2\beta}$ = 2775.39(10) keV [6], isotopic abundance δ = 1.25(6)% [7]) is one of the most suitable nuclei to search for 2β processes with decrease of nuclear charge (see e.g. [8] and references therein). A strong motivation to study the double beta "plus" processes was discussed in [9] where a possibility to distinguish between a right-handed currents admixture and the neutrino mass mechanisms has been considered.

A cadmium tungstate crystal scintillator enriched in $^{106}$Cd to 66% ($^{106}$CdWO$_4$) was developed [10] to search for double beta processes in $^{106}$Cd. The first stage of the experiment realized in the DAMA/R&D set-up at the Gran Sasso underground laboratory was reported in [8].

To increase the experimental sensitivity to the 2β processes with emission of γ quanta, the $^{106}$CdWO$_4$ scintillator was placed inside a low background HPGe detector with four Ge crystals. Here we report first results of the experiment.

## 2 Experiment

The $^{106}$CdWO$_4$ crystal scintillator (mass of 215 g) is viewed through a lead tungstate (PbWO$_4$) crystal light-guide (⌀40 × 83 mm) by 3 inches low radioactive photomultiplier tube (PMT) Hamamatsu R6233MOD (see Fig. 1). The PbWO$_4$ crystal was developed from deeply purified [11] archaeological lead [12]. The detector is installed in an ultra-low background GeMulti HPGe γ spectrometer of the STELLA (SubTErranean Low Level Assay) facilities [13] at the Gran Sasso underground laboratory (LNGS) of the INFN (Italy) on the depth of 3600 m of water equivalent. Four HPGe detectors of the GeMulti set-up are mounted in one cryostat with a well in the centre. The volumes of the HPGe detectors are approximately 225 cm$^3$ each. The

---

typical energy resolution (FWHM) is 2.0 keV for the 1332 keV γ quanta of $^{60}$Co.

An event-by-event data acquisition system is based on two four-channel all digital spectrometers (DGF Pixie-4, XIA, LLC). One device (marked (1) in Fig. 1) is used to provide spectrometric data for the HPGe detectors, while the second Pixie-4 (2) acts as a 14-bit waveform digitizer to acquire signals from the $^{106}$CdWO$_4$ detector at the rate of 18.8 MSPS over a time window 54.8 µs. The second Pixie-4 unit records also trigger signals from the home made unit SST-09, which provides the triggers only if the signal amplitude in the $^{106}$CdWO$_4$ detector exceeds ~ 0.6 MeV to avoid acquisition of a large amount of data caused by the decays of $^{113m}$Cd ($Q_\beta$ = 580 keV) presented in the $^{106}$CdWO$_4$ crystal [8]. The signals from the timing outputs of the HPGe detectors after summing are fed to the third input of the second Pixie-4 digitizer to select off-line coincidence between the $^{106}$CdWO$_4$ and HPGe detectors.

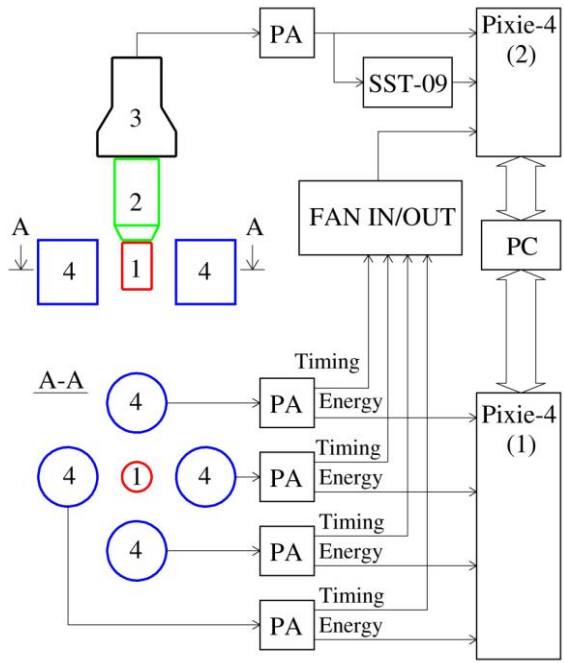

**Figure 1.** Low background $^{106}$CdWO$_4$ crystal scintillator (1) viewed through PbWO$_4$ light-guide (2) by PMT (3). The scintillator is installed between HPGe detectors (4). Schema of the electronic chain: (PA) preamplifiers; (FAN IN/OUT) linear FAN-IN/FAN-OUT; (SST-09) home-made electronic unit to provide triggers for cadmium tungstate scintillation signals; (Pixie-4) four-channel all digital spectrometers; (PC) personal computer.

The detector was calibrated with $^{22}$Na, $^{60}$Co, $^{137}$Cs and $^{228}$Th γ sources. The energy resolution of the detector can be described by the function: FWHM = $(20.4 \times E_\gamma)^{1/2}$, where FWHM and $E_\gamma$ are given in keV.

Energy spectrum and distribution of the start positions of the $^{106}$CdWO$_4$ detector pulses relatively to the HPGe signals accumulated with the $^{22}$Na γ source (see Fig. 2) demonstrate presence of coincidences between the $^{106}$CdWO$_4$ and HPGe detectors under the condition that the energy of events in the HPGe detectors is equal to 511 keV (energy of annihilation gamma quanta), while practically there is no coincidence in the data accumulated with $^{137}$Cs. The measured data is in agreement with the distributions simulated by the EGS4 code [14] (Fig. 2).

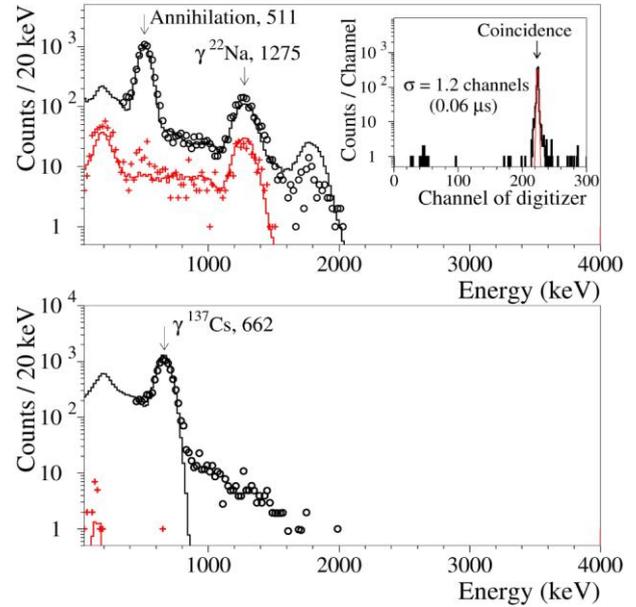

**Figure 2.** Energy spectra of $^{22}$Na (upper figure) and $^{137}$Cs (lower figure) γ sources accumulated by the $^{106}$CdWO$_4$ detector: with no coincidence (circles), and in coincidence with energy 511 keV in the HPGe detector (crosses). The data simulated by using the EGS4 Monte Carlo code are drawn by solid lines. (Inset) Distribution of the $^{106}$CdWO$_4$ detector pulses start positions relatively to the HPGe signals with the energy 511 keV accumulated with $^{22}$Na source (the time shift of the $^{106}$CdWO$_4$ signals of ≈ 220 channels is due to the tuning of the digitizer to provide baseline data).

## 3 Results and discussion

### 3.1 Low background measurements

The mean-time pulse-shape discrimination method (see [8]) was used to discriminate γ(β) events from α events caused by internal contamination of the crystal by uranium and thorium. The energy spectrum of the γ(β) events accumulated in the set-up over 3233 h is shown in Fig. 3. The data confirmed the assumption about surface contamination of the $^{106}$CdWO$_4$ crystal by $^{207}$Bi [8]. The γ peaks of $^{207}$Bi disappeared thanks to the cleaning of the scintillator by potassium free detergent and ultra-pure nitric acid.

The spectrum was fitted by the model built from the energy distributions simulated by EGS4 code. The model includes radioactive contamination of the $^{106}$CdWO$_4$ crystal [8], PMT [15], PbWO$_4$ light-guide, copper shield, aluminium wall of the cryostat. The main components of the background are shown in Fig. 3.

Energy spectra accumulated by the HPGe detector are presented in Fig. 4. The counting rate of the HPGe detector with the $^{106}$CdWO$_4$ counter inside exceeds slightly the background counting rate. Some excess on the level of (30–170)% (depending on the energy of gamma quanta) is observed in the peaks of $^{214}$Bi and $^{214}$Pb (daughters of $^{226}$Ra from $^{238}$U family). Simulation of the background is in progress with an aim to identify materials of the $^{106}$CdWO$_4$ detector contributing to the counting rate excess.

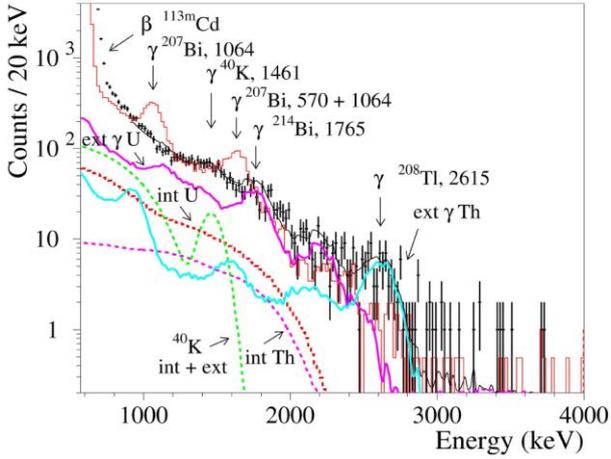

**Figure 3.** The energy spectrum of the β(γ) events accumulated over 3233 h in the low background set-up with the $^{106}$CdWO$_4$ crystal scintillator (points) together with the background model (black continuous superimposed line). The main components of the background are shown: the distributions of internal and external $^{40}$K, internal $^{228}$Th and $^{238}$U, and the contribution from the external γ quanta from U and Th contamination of the set-up in these experimental conditions. The energy spectrum of the β(γ) background accumulated with the $^{106}$CdWO$_4$ crystal scintillator over 6590 h [8] (normalised to 3233 h) is shown by solid histogram. Peaks of $^{207}$Bi disappeared after the cleaning of the $^{106}$CdWO$_4$ crystal.

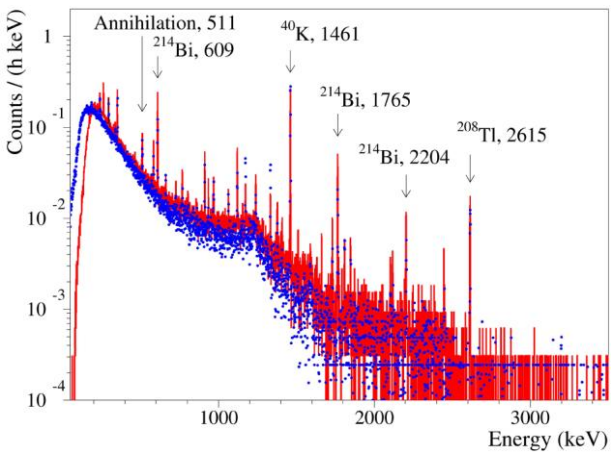

**Figure 4.** Energy spectra accumulated by the low background HPGe γ spectrometer with the $^{106}$CdWO$_4$ counter inside over 3233 h (solid red histogram) and without it over 4102 h (blue dots). Energies of γ lines are in keV.

The counting rate of the $^{106}$CdWO$_4$ detector is substantially suppressed in coincidence with events in the HPGe detector with energy 511 keV (see Fig. 5). The decrease of the background was confirmed also by the Monte Carlo simulation. A counting rate in the coincidence data presented in Fig. 5 (53 counts in the energy interval 50 – 3000 keV) is in agreement with the calculated background (53 events) using the parameters of the fit of the $^{106}$CdWO$_4$ detector background without coincidence.

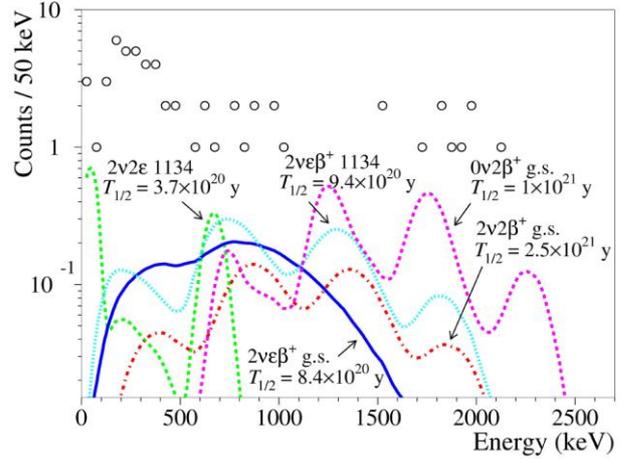

**Figure 5.** Background energy spectrum of the $^{106}$CdWO$_4$ detector in coincidence with 511 keV annihilation γ quanta in the HPGe detectors accumulated over 3233 h (circles) together with the simulated distributions of double beta processes in $^{106}$Cd excluded at 90% C.L.

### 3.2 Sensitivity to the 2β processes in $^{106}$Cd

The response functions of the $^{106}$CdWO$_4$ detector to the 2β processes in $^{106}$Cd were simulated with the help of the EGS4 code. The distributions without coincidence and in coincidence with 511 keV γ quanta in the HPGe detector are presented in Fig. 6.

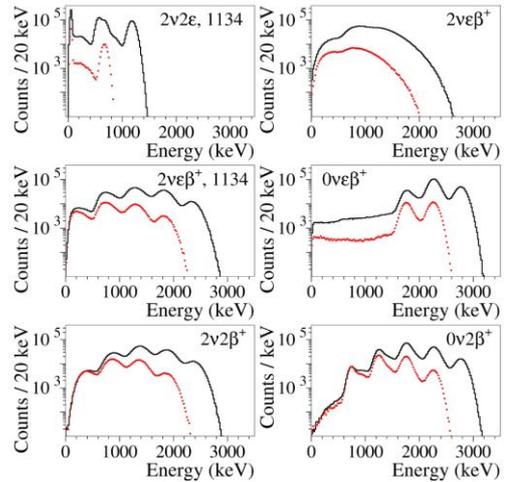

**Figure 6.** Simulated response functions of the $^{106}$CdWO$_4$ detector to 2ε, εβ$^+$, and 2β$^+$ processes in $^{106}$Cd without coincidence (solid histogram) and in coincidence with 511 annihilation γ quanta in the HPGe detector (dotted histogram).

There are no peculiarities in the data accumulated with the $^{106}$CdWO$_4$ detector that could be ascribed to the 2β processes in $^{106}$Cd. Therefore only lower half-life limits can be set by using the formula:

$$\lim T_{1/2} = N \times \eta \times t \times \ln 2 / \lim S,$$

where $N$ is the number of $^{106}$Cd nuclei in the $^{106}$CdWO$_4$ crystal ($2.42 \times 10^{23}$), $\eta$ is the detection efficiency, $t$ is the time of measurements, and $\lim S$ is the number of events of the effect searched for, which can be excluded at a given confidence level (C.L.).

To estimate lim$S$ values for the double beta processes in $^{106}$Cd, the measured number of events in the background spectrum was compared with the expected background, which was estimated by using the result of the fit of the data accumulated by the $^{106}$CdWO$_4$ detector without coincidence (see Fig. 3). For instance, the number of events in the coincidence data in the energy interval 500 – 1200 keV is equal to 13 counts, while the model of background gives 17.6 counts. According to the procedure proposed in [16], we should take 3.7 counts as an effect's limit which can be excluded with 90% C.L. Taking into account the detection efficiency to the two neutrino electron capture with emission of positron in $^{106}$Cd (7.6%), the part of the energy spectrum in the energy interval (67.0%), the selection efficiency of the time and energy cuts used to obtain the coincidence spectrum (totally 99%), one could get a new improved limit on the effect searched for:

$$T_{1/2}^{2\nu\varepsilon\beta^+} \geq 8.4 \times 10^{20} \text{ yr} \quad \text{at 90\% C.L.}$$

Similarly, by using the described procedure, the limits on some other double beta processes in $^{106}$Cd were obtained. Some of the excluded distributions of double beta processes in $^{106}$Cd are presented in Fig. 5. All the half-life limits are summarized in Table 1, where results of the most sensitive previous experiments are given for comparison.

**Table 1.** Half-life limits on 2β processes in $^{106}$Cd to the ground state (g.s.) and to the first $0^+$ 1134 keV excited level of $^{106}$Pd.

| Decay channel, level of $^{106}$Pd (keV) | $T_{1/2}$ limit (yr) at 90% C.L. | |
|---|---|---|
| | Present work | Previous limit |
| $2\nu 2\varepsilon$, $0_1^+$ 1134 | $\geq 3.7 \times 10^{20}$ | $\geq 1.7 \times 10^{20}$ [8] |
| $0\nu 2\varepsilon$, g.s. | $\geq 2.4 \times 10^{19}$ | $\geq 1.0 \times 10^{21}$ [8] |
| $2\nu\varepsilon\beta^+$, g.s. | $\geq 8.4 \times 10^{20}$ | $\geq 4.1 \times 10^{20}$ [17] |
| $2\nu\varepsilon\beta^+$, $0_1^+$ 1134 | $\geq 9.4 \times 10^{20}$ | $\geq 3.7 \times 10^{20}$ [8] |
| $0\nu\varepsilon\beta^+$, g.s. | $\geq 4.3 \times 10^{20}$ | $\geq 2.2 \times 10^{21}$ [8] |
| $2\nu 2\beta^+$, g.s. | $\geq 2.5 \times 10^{21}$ | $\geq 4.3 \times 10^{20}$ [8] |
| $0\nu 2\beta^+$, g.s. | $\geq 1.0 \times 10^{21}$ | $\geq 1.2 \times 10^{21}$ [8] |

## 4 Conclusions

An experiment to search for double beta decay processes in $^{106}$Cd with the help of enriched in $^{106}$Cd (to 66%) low background $^{106}$CdWO$_4$ scintillation detector (215 g) in coincidence with the four crystals HPGe γ spectrometer GeMulti (the volume of each detector is 225 cm$^3$) is in progress at the STELLA facility of the Gran Sasso underground laboratory of INFN (Italy). The sensitivity of the experiment after 3233 h of data taking is on the level of lim $T_{1/2}$ ~ $10^{19}$ – $10^{21}$ yr for the double β processes in $^{106}$Cd with emission of γ quanta. In particular, a new improved half-life limit is set to the two neutrino electron capture with positron emission in $^{106}$Cd as $T_{1/2} \geq 8.4 \times 10^{20}$ yr (at 90% C.L.). We hope to improve further the sensitivity of the experiment particularly for this channel of the decay to the level of theoretical predictions ($T_{1/2}$ ~ $10^{20}$ – $10^{22}$ yr [9, 18, 19, 20, 21, 22]) by increase of the statistics and by construction of a more precise model of the background.

## Acknowledgment

The authors from the Institute for Nuclear Research (Kyiv, Ukraine) were supported in part by the Space Research Program of the National Academy of Sciences of Ukraine. D.M.Ch., V.M.M. and D.V.P. have been also supported by a grant of the National Academy of Sciences of Ukraine for young scientists (reg. no. 0113U005378).